\documentclass[aps,preprint,prd ,epsfig]{revtex4}
\usepackage{amssymb}
\usepackage{amsmath}
\usepackage{graphicx}
\usepackage{epsfig}
\usepackage{ccaption}
\usepackage{color}
\usepackage{fancyhdr}
\usepackage{textcomp}
\usepackage{tikz}
\newcommand{\be}{\begin{equation}}
\newcommand{\ee}{\end{equation}}
\newcommand{\bea}{\begin{eqnarray}}
\newcommand{\eea}{\end{eqnarray}}
\newcommand{\bes}{\begin{subequations}}
\newcommand{\ees}{\end{subequations}}

\newcommand{\bc}{\begin{center}}
\newcommand{\ec}{\end{center}}
\usepackage{wasysym}

\begin{document}
\title{  Neutrino masses, cosmological inflation and dark matter in a $U(1)_{B-L}$ model with type II seesaw mechanism.}
\author{J. G. Rodrigues, A. C. O. Santos, J. G. Ferreira Jr,  C. A. de S. Pires}
\affiliation{{ Departamento de F\'{\i}sica, Universidade Federal da Para\'\i ba, Caixa Postal 5008, 58051-970,
Jo\~ao Pessoa, PB, Brazil}}

\begin{abstract}
In this work we implement the type II seesaw mechanism into the framework of the   $U(1)_{B-L}$ gauge model. As main gain, the right-handed neutrinos of the model get free to play the role of the dark matter of the universe. As side effect, the model realizes Higgs inflation without problem with loss of unitarity.
\end{abstract}
 \maketitle
\date{\today}

\section{Introduction}
Experimental observation of neutrino oscillation coming from the sum and from the atmosphere surprisingly revealed that neutrinos are very light particles.\cite{Patrignani:2016xqp}. From the theoretical point of view, seesaw mechanism is the most popular way of generating tiny neutrino  masses.

 The observation of galaxies rotation curves\cite{Faber:1979pp}, cluster collisions\cite{Clowe:2006eq} and the precise measurements of the thermal anisotropy of the cosmic microwave background\cite{Ade:2015xua}    suggest the existence of dark matter (DM) permeating our universe. Recent results from PLANCK satellite indicates that 26.7\% of the matter of the universe is in the form of non-luminous matter \cite{Ade:2015xua}.  The most popular DM candidate is a weakly interactive massive particle (WIMP)\cite{Jungman:1995df, Bertone:2004pz}. WIMPs can be any kind of particle since they are neutral, stable (or sufficiently long lived) and have mass in the range from few GeV's up to some TeV's. 

 Cosmological inflation is considered the best theory  for explaining homogeneity, flatness and isotropy of the universe as required by hot big bang\cite{Guth:1980zm,Linde:1981mu,Albrecht:1982wi}. Experiments in cosmology, as WMAP7 and PLACK15 \cite{Bennett:2012zja,Ade:2015lrj}, entered in an era of precision which allow us to probe  proposal of physics scenario that try to explain the very early  universe. Single-field slow-roll models of inflation coupled non-minimally to gravity  appear to be an interesting scenario for inflation\cite{Makino:1991sg,Bezrukov:2007ep} since it  connects inflation to particle physics at low energy scale\cite{Bezrukov:2007ep}.

Although the standard model (SM) of particle physics is a very successful theory, its framework does not accommodate any one of the three issues discussed above. In other words, nonzero neutrino mass, dark matter and inflation require extensions of the SM. 

In this work we show that  the $U(1)_{B-L}$ gauge model is capable of accomplishing all these three issues in a very attractive way by simply adding a scalar triplet to its canonical scalar sector \cite{Mohapatra:1980qe,Appelquist:2002mw,Basso:2008iv, Khalil:2010iu,Okada:2011en}.  As nice results, we have that small neutrino masses is achieved through the type II seesaw mechanism, which is triggered by  the spontaneous breaking of the B-L symmetry, while the dark matter content of the universe is composed  by the lightest right-handed neutrino of the model and,  by allowing non-minimal coupling of gravity with scalars, we show that the model perform inflation without loss of unitarity. 

This work is organized as follow: in Section II, we describe the main properties of the B-L model. Section III is devoted to cosmological inflation. In Section IV we describe our calculation of the dark matter candidate, and Section V contains our conclusions.

\section{The B-L Model with scalar triplet}
\subsection{The seesaw mechanism}
Baryon number (B) and lepton number (L) are accidental anomalous symmetries of the SM.  However it is well known that only some specific linear combinations of these symmetries can be free from anomalies\cite{Mohapatra:1980qe,He:1990pn,He:1991qd,Ma:1997nq}. Among them the most developed one is the B-L symmetry\cite{Mohapatra:1980qe,Appelquist:2002mw,Basso:2008iv, Khalil:2010iu} which is involved in several physics scenarios such as GUT\cite{Goh:2004fy}, seesaw mechanism\cite{Minkowski:1977sc,Yanagida:1979as,GellMann:1980vs,Mohapatra:1979ia} and baryogenesis\cite{Fukugita:1986hr}.  This symmetry engenders the simplest gauge extension of the SM, namely, the \textit{B-L model} which is based on the gauge group $SU(3)_C \times SU(2)_L \times U(1)_Y \times U(1)_{B-L}$. In this work we consider an extension of the B-L model with the scalar sector being composed by a scalar triplet in addition to the doublet and scalar singlet of the minimal version. In this way the particle content of the model involves the standard particle content augmented by three RHNs, $N_i \,\,,\,\,i=1,2,3$, one scalar singlet, $S$, and one scalar triplet, 
\begin{equation}
\Delta\equiv \left(\begin{array}{cc}
\frac{\Delta^{+}}{\sqrt{2}} & \Delta^{++} \\ 
\Delta^{0} & \frac{-\Delta^{+}}{\sqrt{2}}
\end{array} \right). 
\end{equation}
As far as we know, this is the first time the triplet $\Delta$ appears composing the scalar sector of the B-L model. Moreover, we impose the model to be invariant by a $Z_2$ discrete symmetry with the RHNs transforming as $N_i \rightarrow -N_i$ while the rest of the particle content of the model transforms trivially by $Z_2$.

With these features, the Yukawa interactions of interest is composed by the terms
\begin{equation}
{\cal L}_{B-L} \supset  Y_\nu\overline{f^C}i\sigma^2 \Delta f + \frac{1}{2}Y_N\overline{N^c} N S + h.c.,
\label{BL}
\end{equation}
where  $f=(\nu \,\,\,\,e)_L^T$. Perceive that both neutrinos gain masses when $\Delta^0$ and $S$  develop a nonzero vacuum expectation value ($v_\Delta$ and $v_S$). This yield he following expressions to the masses of these neutrinos
\begin{equation}
m_\nu= \frac{Y_\nu v_\Delta}{\sqrt{2}}\,\,\,\,,\,\,\,\,  m_{\nu_R}=\frac{Y_N v_S}{\sqrt{2}}.
\label{numasses}
\end{equation}
Observe that there is no mixing mass terms involving $N$ and $\nu_L$ like in the type I seesaw mechanism. The energy scale of the neutrino masses is defined by $v_\Delta$ and $v_S$.  Thus, small masses for the standard neutrinos requires a tiny $v_\Delta$. We are going to show that on fixing $v_h$ and $v_S$ we may obtain $v_\Delta$ around eV scale \`a la type II seesaw mechanism  \cite{PhysRevD.22.2227,PhysRevD.23.165,Arhrib:2011uy}.  For this we must develop the potential of the model which is invariant by the B-L symmetry and involves the following terms
\bea
V(H,\Delta,S) = && \mu^2_h H^\dagger H + \lambda_h (H^\dagger H)^2 + \mu^2_s S^\dagger S + \lambda_s (S^\dagger S)^2 \nonumber \\
&& + \mu^2_\Delta Tr(\Delta^\dagger \Delta) + \lambda_\Delta Tr[(\Delta^\dagger \Delta)^2] +\lambda^\prime_\Delta Tr[(\Delta^\dagger \Delta)]^2 \nonumber \\
&& + \lambda_1 S^\dagger SH^\dagger H + \lambda_2 H^\dagger\Delta \Delta^\dagger H + \lambda_3 Tr(\Delta^\dagger \Delta)H^\dagger H \nonumber \\
&& + \lambda_4 S^\dagger STr(\Delta^\dagger \Delta) + (k Hi\sigma^2\Delta^\dagger  HS + h.c.).
\eea
where $S$ is the scalar singlet by the standard symmetry  and $H=(h^+ \,\,\,h^0)^T$ is the standard Higgs doublet. We assume that all neutral scalars develop vev different from zero. To obtain the set of conditions that guarantee such potential have a global minimum, we must  shift the neutral scalar fields in the conventional way
\begin{equation}
S,h^0,\Delta^0\rightarrow \frac{1}{\sqrt{2}}(v_{S,h,\Delta}+R_{S,h,\Delta}+iI_{S,h,\Delta}),
\label{shift}
\end{equation}
and then substitute them in the potential above. Doing this we obtain the following set of  minimum condition equations
\bea
&& v_S\left(\mu^2_S + \frac{\lambda_1}{2}v^2_h + \frac{\lambda_4}{2}v^2_\Delta + \lambda_s v^2_S\right) -\frac{k}{2}v^2_h v_\Delta=0, \nonumber \\
&& v_h \left(\mu^2_h + \frac{\lambda_1}{2}v^2_S + \frac{\lambda_2}{2}v^2_\Delta + \frac{\lambda_3}{2}v^2_\Delta + \lambda_h v^2_h - k v_\Delta v_S\right) =0, \nonumber \\
&& v_\Delta \left(\mu^2_\Delta + \frac{\lambda_2}{2}v^2_h + \frac{\lambda_3}{2}v^2_h + \frac{\lambda_4}{2}v^2_S + (\lambda_\Delta + \lambda^{\prime}_\Delta) v^2_\Delta\right) -\frac{k}{2}v^2_h v_S=0.
\label{vinc}
\eea

Remember that $v_\Delta$ modifies softly the $\rho$-parameter in the following way: $\rho=\frac{1+\frac{2v^2_\Delta}{v^2_\phi}}{1+\frac{4v^2_\Delta}{v^2_\phi}}$. The electroweak precision data constraints require the value $\rho=1.00037 \pm 0.00023$\cite{Patrignani:2016xqp}. This implies the following upper bound $v_\Delta < 4$GeV. Consequently, $v_S$ passes to contribute dominantly to the mass of the new neutral gauge boson $Z^{\prime}$ associated to the B-L symmetry which also has a constraint due the LEP experiment\cite{Carena:2004xs} 
\begin{equation}
\frac{m_{Z^{\prime}}}{g_{B-L}} \gtrsim 6.9 \mbox{TeV}.
\label{LEP}
\end{equation}

Perceive that the third relation in Eq. (\ref{vinc}) provides
\begin{equation}
v_\Delta \approx \frac{k}{2}\frac{v^2_h v_S}{\mu^2_\Delta}.
\label{vev}
\end{equation}
The role of the type II seesaw mechanism is to provide tiny vevs. In the canonical case, where $v_\Delta=\frac{v^2_h}{\mu}$,  tiny $v_\Delta$ is a consequence of the explicit violation of the lepton number which must happens at GUT scale ($\mu=10^{14}$ GeV). Observe that in our case   $v_\Delta$ get suppressed by the quadractic term $\mu^2_\Delta$. This allows we have a seesaw mechanism occuring  in an intermediate energy scale, as we see below. 

Remember that $v_h$ is the standard vev whose value is $247$ GeV while $v_S$ define the mass of the neutral gauge bosons $Z^{\prime}$ and its  value must lie around few TeVs. Here we take  $v_S \sim 10$ TeV. This provides  $v^2_h v_S \sim 10^9$ GeV$^3$. Consequently,  $v_\Delta \sim$ eV requires  $\mu_\Delta \sim 10^9$ GeV. As conclusion, we have that  type II seesaw mechanism  engendered  by the spontaneous breaking of the lepton number is associated to a new physics in the form of scalar triplet with mass around $10^9$ GeV.  Such energy scale is too high to be probed at the LHC. However such regime of energy  may give sizable contributions in flavor physics and then be probed through rare lepton decays. This point will be discussed elsewhere.

\subsection{Spectrum of scalars}
\begin{figure}[!h]
  \centering
 \includegraphics[width=0.5\textwidth]{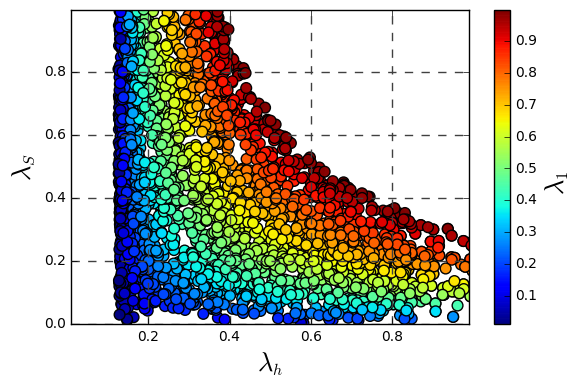} 
\caption{Possible values of the quartic couplings that yield 125 GeV Higgs mass.}
  \label{higgspar}
  \end{figure}

Before go further, it makes necessary to discuss briefly  the scalar sector of the model. Let us first focus on the CP-even sector. In the basis $(R_S,R_h,R_\Delta)$ we have the following mass matrix,
\be
M^2_R= \left(\begin{array}{ccc}
\frac{k}{2}\frac{v_\Delta v^2_h}{v_s}+2\lambda_S v^2_s & -kv_h v_\Delta + \lambda_1 v_s v_h & -\frac{k}{2} v^2_h + \lambda_4 v_s v_\Delta\\ 
-kv_h v_\Delta + \lambda_1 v_s v_h & 2\lambda_h v^2_h & -k v_s v_h + (\lambda_2 + \lambda_3) v_h v_\Delta \\
-\frac{k}{2} v^2_h + \lambda_4 v_s v_\Delta & -k v_s v_h + (\lambda_2 + \lambda_3) v_h v_\Delta & \frac{k}{2}\frac{v_s v^2_h}{v_\Delta} + 2(\lambda_\Delta + \lambda^\prime_\Delta)v^2_\Delta
\end{array} \right).\label{scalmass}
\ee
Note that for values of the vevs  indicated above, the scalar $R_\Delta$ get very heavy, with $m^2_\Delta \sim \frac{k}{2}\frac{v_S v^2_h}{v_\Delta}$, which implies that it decouples from the other ones. The other two quadratic masses are
\begin{eqnarray}
m^2_h&\simeq & 2\lambda_h v^2_h -\frac{1}{2}\frac{\lambda_1^2}{\lambda_S}v^2_h,\nonumber \\
 m^2_H &\simeq & 2\lambda_S v^2_S +\frac{1}{2}\frac{\lambda_1^2}{\lambda_S}v^2_h,
 \label{higgsmass}
\end{eqnarray}
where $m_h$ stands for the standard Higgs boson with the allowed parameter space showed in FIG. \ref{higgspar}

The respective eigenvectors are
\begin{eqnarray}
h&\simeq &R_h-\frac{\lambda_1}{2\lambda_S}\frac{v_h}{v_s}R_S,\nonumber \\
H&\simeq &R_S+\frac{\lambda_1}{2\lambda_S}\frac{v_h}{v_s}R_h.
\label{eigenvectorsR}
\end{eqnarray}

For the CP-odd scalars, we have the mass matrix in the basis $(I_S,I_h,I_\Delta)$,

\be
M^2_I= \left(\begin{array}{ccc}
\frac{k}{2}\frac{v_\Delta v^2_h}{v_s} & kv_h v_\Delta & -\frac{k}{2} v^2_h \\ 
kv_h v_\Delta & 2k v_s v_\Delta & -k v_s v_h \\
-\frac{k}{2} v^2_h  & -k v_s v_h & \frac{k}{2}\frac{v_s v^2_h}{v_\Delta}
\end{array} \right).\label{pseumass}
\ee
The mass matrix in Eq. (\ref{pseumass}) can be diagonalized providing one massive state $A^0$ with mass,
\be
m^2_A=\frac{k}{2}\left(\frac{v_\Delta v^2_h}{v_s} + \frac{v_s v^2_h}{v_\Delta} + 4v_s v_\Delta \right),
\ee
and two goldstone bosons $G^1$ and $G^2$, absorbed as the longitudinal components of the $Z$ and $Z^{\prime}$ gauge bosons. The eigenvectors for the cp-odd scalars are,
\begin{eqnarray}
G^1 &\simeq & I_S + \frac{v_\Delta}{v_s} I_\Delta,\nonumber \\
G^2 &\simeq & I_h + \frac{v_h}{2v_s} I_S,\nonumber \\
A^0 &\simeq & I_\Delta - \frac{2v_\Delta}{v_h} I_h.
\end{eqnarray}

The charged scalars, given in the basis $(h^+,\Delta^+)$, have the mass matrix
\be
M^2_+= \left(\begin{array}{cc}
k v_s v_\Delta - \frac{\lambda_2}{2}v^2_\Delta & \frac{\lambda _2}{2\sqrt{2}}v_h v_\Delta -\frac{k}{\sqrt{2}}v_s v_h \\ 
\frac{\lambda _2}{2\sqrt{2}}v_h v_\Delta -\frac{k}{\sqrt{2}}v_s v_h & \frac{k}{2} \frac{v_s v^2_h}{v_\Delta} - \frac{\lambda_2}{2}v^2_h \end{array}\right) .\label{charmass}
\ee
Again, diagonalizing this matrix gives us two goldstone bosons $G^\pm$, responsible by the longitudinal parts of the $W^\pm$ standard gauge bosons. The other two degrees of freedom give us a massive states $H^\pm$ with mass
\be
m^2_{H^\pm} = \left( \frac{v_\Delta}{2} + \frac{v^2_h}{4v_\Delta} \right)\left( 2k v_s - \lambda_2 v_\Delta \right).
\ee
The respective eigenvectors are,
\begin{eqnarray}
G^\pm &\simeq & h^{\pm} + \frac{\sqrt{2}v_\Delta}{v_h} \Delta^{\pm},\nonumber \\
H^{\pm} &\simeq & \Delta^{\pm} - \frac{\sqrt{2}v_\Delta}{v_h}h^{\pm}.
\end{eqnarray}

Finally the mass of the doubly charged scalars $\Delta^{\pm \pm}$ are expressed as,
\be
m^2_{\Delta^{\pm \pm}} = \frac{k v_s v^2_h v_\Delta - \lambda_2 v^2_h v^2_\Delta -2\lambda_\Delta v^4_\Delta}{2v^2_\Delta}.
\ee

Once symmetries are broken and the gauge bosons absorb the goldstone bosons  as longitudinal component, we have that the standard charged bosons 
get a contribution from the triplet vev, $m^2_w=\frac{g^2}{4}\left(v^2_h + 2v^2_\Delta \right)$, while  the neutral gauge bosons get mixed with $Z^{\prime}$ in the following way
\be
M^2_g= \left(\begin{array}{cc}
 \frac{g^2+g^{\prime 2}}{4}(v^2_h+4v_\Delta^2) & -g_{B-L}\sqrt{g^2+g^{\prime 2}}v_\Delta^2 \\
-g_{B-L}\sqrt{g^2+g^{\prime 2}}v_\Delta^2 & g^2_{B-L}(2v_S+v_\Delta^2)
\end{array} \right).\label{gaumass}
\ee
Recording the vev hierarchy discussed here ($v_S>v_h\gg v_\Delta$), the mixing between the gauge bosons are very small, therefore, they decouple and we have the masses,
\be
\quad M^2_Z\approx \frac{(g^2+g^{\prime 2})( v^2_h + 4 v_\Delta^2)}{4}, \quad M^2_{Z^{\prime}}\approx 2g^2_{B-L}(v^2_S+ \frac{v_\Delta^2}{2}).
\ee

Observe that we have a B-L model which has as new ingredients scalars in the triplet and singlet forms and neutrinos with right-handed chiralities. Let us resume the role played by these new ingredients. The singlet $S$ is responsible by the spontaneous breaking of the B-L symmetry and then define the mass of the $Z^{\prime}$. The triplet $\Delta$ is responsible by the type II seesaw mechanism that generate small masses for the standard neutrinos. The right-handed neutrinos is responsible by the cancellation of anomalies. It would be interesting to find new roles for these ingredients.

We argue here, and check below, that the right-handed neutrinos may be the dark matter component of the universe since the $Z_2$ symmetry  protect them of decaying in lighter particles. In the last section we assume that the lightest right-handed neutrinos is the dark matter of the universe and then calculate its abundance and investigate the ways of detecting it. 

We also argue here that once $\Delta^0$ has mass around $10^9$ GeV, it could be possible that it would come to be the inflaton and then drives inflation. We show in the next section that this is possible when we assume non-minimal coupling of $\Delta$ with gravity.

\section{Cosmological Inflation}

For successful inflation we have to evoke non-minimal coupling of the inflaton with gravity \cite{Salopek:1988qh,Fakir:1990eg,Makino:1991sg}. This has been done in an extensive list of models where the Higgs field \cite{Bezrukov:2007ep,Barvinsky:2008ia,GarciaBellido:2008ab,Bezrukov:2014bra,Lee:2018esk} or a standard model singlet extension assume the role of inflaton \cite{Lerner:2009xg,Okada:2010jf,Fairbairn:2014zta}. However, such models lead to a troublesome behavior in the low scale phenomenology. For the case of Higgs Inflation, the measured Higgs mass pushes the non-minimal coupling to high values ($\xi \sim 10^5$), causing unitarity problems at inflationary scale \cite{Burgess:2009ea,Barbon:2009ya,Burgess:2010zq,Hertzberg:2010dc}. On the other hand, the singlet scenario is also problematic. Although one could manage to built a unitarity safe singlet inflation, this would produce a very light inflaton, jeopardizing the reheating period of universe \cite{Ferreira:2017ynu}.

With the scalar triplet things is different  thanks to the mass structure of the CP-even mass matrix presented above whose diagonalization implies that  $m^2_{\Delta}\sim \frac{k}{2}\frac{v_S v^2_h}{v_\Delta}$. Note that the inflaton's mass is independent of any inflationary parameter ($\lambda_\Delta, \lambda^\prime_\Delta,\xi$). Such configuration yields a unitarity safe inflationary model which does not put in risk the transition to the standard evolution of the universe.

As argued above, $\Delta^0$ may be the inflaton and then drives inflation.  For previous works of inflation  conducted by $\Delta^0$, see \cite{Chen:2010uc,Arina:2012fb,Ferreira:2017ynu}. For sake of simplicity, we assume that $\Delta^0$ provides the dominant coupling.  Thus, the relevant terms in 
the potential become the quartic one $V(\Delta^0)=\frac{\lambda_\Delta +\lambda^\prime_\Delta}{4} {\Delta^0}^4$.  In this case we have that the lagrangian of the model must involves the following  terms in the Jordan frame,

\be
 {\cal L} \supset \frac{1}{2} (\partial_\mu \Delta^0)^{\dagger}(\partial^\mu \Delta^0)-\frac{M_P^2R}{2}-\frac{1}{2}\xi {\Delta^0}^2 R -V(\Delta^0).
 \label{Ljordan}
 \ee

 In order to calculate the parameters related to inflation we must recover the canonical 
Einstein-Hilbert gravity. This process is called conformal transformation and can be understood by two steps. First we re-scale 
the metric $\tilde{g}_{\alpha \beta} = \Omega^2 g_{\alpha \beta}$. In doing so the coupling to gravity vanish but the inflaton acquire 
a non-canonical kinect term. The process is finished by transforming the inflaton field to a form with canonical kinect energy. 
Such transformation involves the relations \cite{Birrell:1982ix,Accioly:1993kc}

\bea
\tilde g_{\mu \nu}&&=\Omega^2g_{\mu \nu}\,\,\,\,\,\,\mbox{where}\nonumber \,\,\,\,\,  \Omega^2= 1+\frac{\xi {\Delta^0}^2}{M^2_P},\nonumber \\
&& \frac{d\chi}{d\Delta^0}=\sqrt{\frac{\Omega^2 +6\xi^2 {\Delta^0}^2/M_P^2}{\Omega^4}}.
\label{conformaltransf}
\eea
The lagrangian in Einstein frame is given by
\be
 {\cal L} \supset -\frac{M^2_{P} \tilde R}{2}+\frac{1}{2} (\partial_\mu \chi)^{\dagger}(\partial^\mu \chi)-U(\chi)\,,
 \label{LEinstein}
\ee
where $U(\chi)=\frac{1}{\Omega^4}V(\Delta[\chi])$. There is some discussion about which frame is the physical one \cite{Faraoni:1998qx}, however 
both frames agrees in the regime of low energy. 


Inflation occurs whenever the field $\chi$, or equivalently $\Delta^0$, rolls slowly in direction to the minimum of the potencial. The slow-roll parameters can be writen as,
\begin{eqnarray}
 &\epsilon & = \frac{M^2_{P}}{2}\left(\frac{ V^{\prime}}{ V \chi^{\prime}}\right)^2, \quad \quad
 \eta  = M^2_{P}\left( \frac{V^{\prime \prime}}{V \chi^{\prime}} - \frac{V^{\prime} \chi^{\prime \prime}}
 {V {\chi^{\prime}}^3}\right),
 \label{slowparameters}
\end{eqnarray}
where $^\prime$ indicate derivative with respect to $\Delta^0$. 
Inflation starts when
$\epsilon,\eta \ll 1$ and stop when $\epsilon,\eta = 1$. In the slow-roll regime we can write the spectral index and the 
scalar-to-tensor ratio as \cite{Liddle:2000cg},
\be
 n_S=1-6\epsilon+2\eta, \quad \quad \quad r=16\epsilon.
 \label{PParameters}
\ee
Planck2015 have measured $n_S=0.9644 \pm 0.0049$ and gave the bound $r<0.149$ for a pivot scale 
equivalent to $k=0.05$~Mpc$^{-1}$\cite{Ade:2015lrj}. Any inflationary model that intend to be realistic must recover these values.

The amount of ``visible" inflation is quantified by the number of e-folds,
\begin{equation}
N = -\frac{1}{M^2_P}\int_{(\Delta^0)_i}^{(\Delta^0)_f}\frac{V(\Delta^0)}{V^\prime(\Delta^0)}\left(\frac{d\chi}{d\Delta^0}\right)^2d\Delta^0.\label{efold}
\end{equation}
For a $\phi^4$ theory the universe is radiative dominated during the reheat, so it is plausible to affirm 
that in our model the number of e-folds should be around 60 \cite{Liddle:2003as,Ballesteros:2016xej}. Using $N=60$ we can solve Eq. (\ref{efold}) for $(\Delta^0)_i$. Substituting it in Eqs. (\ref{PParameters}) we obtain  the spectral index and tensor-to-scalar ratio. Before going to this point, we must consider the effect of radiative corrections for the scalar potential. This makes necessary because such corrections could modify the shape of the inflationary potential, altering the predicted values for inflationary parameters.

 As we are dealing with an inflaton that is triplet under $SU(2)_L$, such corrections will involve the standard gauge couplings $g$, $g^{\prime}$ and $g_{B-L}$.   Here we consider one-loop radiative correction in Jordan frame, as done in \cite{Barvinsky:2008ia,GarciaBellido:2008ab,Lerner:2009xg,Okada:2010jf}. The complete potential  involving the relevant terms for inflation is given by
\be
V=\left( \frac{\lambda_{\Delta} + \lambda^{\prime}_{\Delta}}{4} + 
\frac{ 3(g^4+{g^{\prime}}^4+{g_{B-L}}^4) - 4Y^4_\nu + \sum_{i} \lambda^2_i}{32\pi^2}\ln{\frac{\Delta^0}{M_P}}\right){\Delta^0}^4, \label{radi}
\ee
where $M_P$ is chosen for renormalization scale and $i$ runs for the scalar contributions ($\lambda_{2}$, $\lambda_{3}$, $\lambda_{4}$, $\lambda_{\Delta}$ and $\lambda^{\prime}_{\Delta}$).  At grand unification scale the standard gauge couplings are evaluated at $g^2 \approx g^{\prime 2}\approx 0.3$ \cite{GarciaBellido:2008ab}. Here we assume $g_{B-L}<g, g^{\prime}$ at any energy scale. 
For simplicity, we will also assume $\lambda_i<<1$. In view of this, the dominant terms in the potential are
\be
V=\left( \frac{\lambda_{\Delta} + \lambda^{\prime}_{\Delta}}{4} + 
\frac{ 3(g^4+{g^{\prime}}^4)- 4Y^4_\nu}{32\pi^2}\ln{\frac{\Delta^0}{M_P}}\right){\Delta^0}^4.
\ee

In the case of the standard Higgs inflation\cite{Bezrukov:2007ep}, as the standard quartic coupling , $\lambda$, is already fixed at $0.6$, the tree level contribution in the effective potential of the inflaton gets dominant over the radiative corrections, and then cosmological constraints imply  $\xi \sim 10^4$ which yield loss of unitarity  at $\Lambda_U = \frac{M_P}{\xi} \sim 10^{14}$ GeV \cite{Burgess:2009ea,Barbon:2009ya}.

Differently from the standard case,  $\lambda_\Delta$ and $\lambda_{\Delta^{\prime}}$ are free parameters. 
Taking into account that the standard scenarios lead to a quartic coupling of order $10^{-13}$, it is very likely that the magnitude of the radiative terms overcomes the tree level contribution in the potential. For the radiative dominance case, the inflaton potential at inflationary time reads,
\be
V=\frac{ 3(g^4+{g^{\prime}}^4- 4Y^4_\nu)}{32\pi^2}\ln{\frac{\Delta^0}{M_P}}{\Delta^0}^4. \label{InfPot}
\ee
Even though the first order terms dominate over the tree level potential it is not necessarily true for higher order contributions. As we are dealing with perturbative parameters, higher loop diagrams are subdominant and we can approximate the inflationary potential as given in (\ref{InfPot}).

\begin{figure}[!htb]
\centering
\includegraphics[scale=0.4]{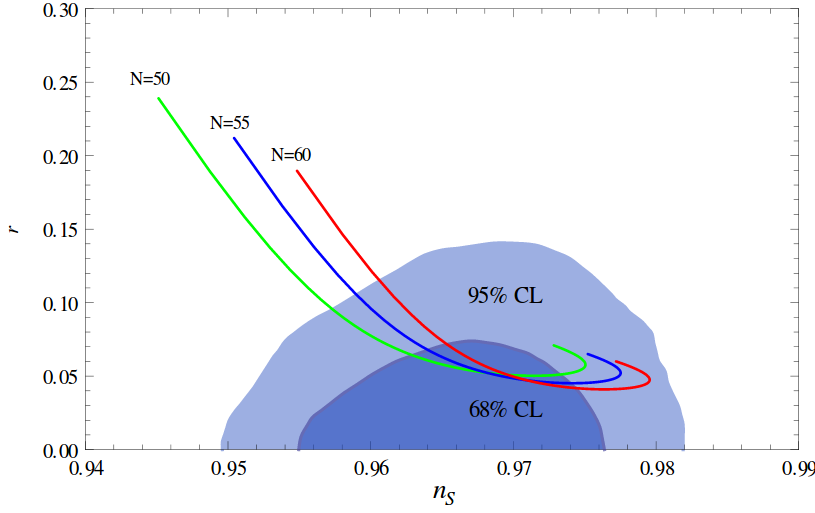}
\caption{$r$ vs $n_S$ for $N = 50$ (green curve), $55$ (blue curve) e $60$ (red curve). The blue contours show the $68\%$ and $95\%$ CL regions observed by Planck2015 \cite{Ade:2015lrj}. $\xi$ is evaluated in the range 
$10^{-3}\leq \xi \leq 1$ for each point in the figure.
}
\label{InfParameters}
\end{figure}

Finally, we can use expressions in (\ref{PParameters}) to calculate the predictions of our model for $n_S$ and $r$ parameters. In fig. \ref{InfParameters} we present our results in the $n_S \times r$ plane. In particular, for $\xi=0.01$ and $N=60$, we obtain $n_S\simeq 0.9666$ and $r\simeq 0.0640$, well inside the $68\%$ CL contour of the most stringent data set of Planck2015 ($Planck \,\,\,\,\, TT+TE+EE+LowP$) \cite{Ade:2015lrj}. Also, we have a unitarily safe model as inflation takes place for $7.92 \times 10^{18} \leq \Delta^0 \leq 6.06 \times 10^{19}$ GeV, well bellow the unitarity scale $\Lambda_U = \frac{M_P}{\xi}\sim 10^{20}$ GeV.

Another important result is that our model predicts sizable tensor-to-scalar ratio, $r \gtrsim 0.4$. This suggest observable primordial gravity waves. 

However, the most striking result comes from the amplitude of scalar perturbations, $A_S$, fixed by Planck2015 to be $A_S \simeq 2.21 \times 10^{-9}$, for pivot scale $k_*=0.05$ Mpc$^{-1}$. The observed value for $A_S$ constrain the neutrino's Yukawa coupling in the inflationary potential to be $Y_\nu(M_P) \simeq 0.460578$. This is a remarkable result, as the Yukawa coupling is associated with the standard neutrinos masses through eq. (\ref{numasses}). To evaluate the impact of this bound at neutrino's masses we should use renormalization group equations to obtain $Y^i_\nu$ at electroweak scale and consider the full structure of the PMNS matrix \cite{Tanabashi:2018oca}. We shall consider this more complete analyses in a future paper.

After the inflationary period, the inflaton oscillates around its vev giving rise to the reheating phase \cite{Abbott:1982hn,Linde:2005ht,Allahverdi:2010xz}. Due to its mass structure, the inflaton is massive enough to decay in pairs of gauge bosons, neutrinos or even the Higgs field. This could result in a reheting temperatures as high as $10^{9}$ GeV \cite{Ferreira:2016uao}. Even before the inflaton settle at its vev, non-perturbative effects could take place producing gauge bosons \cite{GarciaBellido:2008ab,Ema:2016dny}. In this case things are significantly more complicated and numerical study in lattice is made necessary.


%


\label{sectionIII}
\section{Dark Matter }

In our model the three RHNs transform non trivially by the $Z_2$ symmetry. Consequently the model does not perform the type I seesaw mechanism. As we saw above, neutrino mass is achieved through an adapted type II seesaw mechanism. In view of this the role played by the RHNs  is to cancel gauge anomalies and, as consequence of the $Z_2$ symmetry,   provide the dark matter content of the universe in the form of WIMP. However, differently from the minimal B-L model\cite{Okada:2011en}, here the three RHNs may be stable particles since we chose $Y_N$ diagonal in Eq. (\ref{BL}). In other words, the model may accommodate multiple DM candidates.

\begin{figure}[!ht]
\centering
\includegraphics[width=0.9\textwidth ]{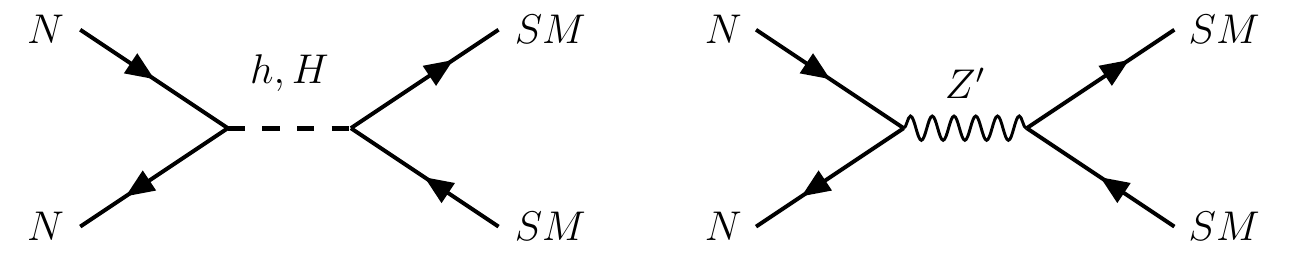}
\caption{ The main contributions to the DM relic Abundance. The SM contributions stands for fermions, Higgs and vector bosons. }
\label{DMRelicYdiagrams0}
\end{figure}

Although the three RHNs are potentially DM candidate, for simplicity reasons we just  consider that  the lightest one, which we call N, is sufficient to provide the correct relic abundance of DM of the universe in the form of WIMP. This means  that $N$ was in thermal equilibrium with the SM particles in the early universe. Then, as far as the universe expands and cools the thermal equilibrium is lost causing the freeze out of the abundance of $N$. This happens when $N$ annihilation rate, whose main contributions are displayed in FIG. \ref{DMRelicYdiagrams0},  becomes roughly smaller then the expansion rate of the universe. In this case the relic abundance of $N$ is obtained by  evaluating 
 the Boltzmann equation for the number density  $n_N$,
\begin{equation}
 \frac{d n_N}{dt}+3H n_N=-\langle\sigma v\rangle(n_N^2 - n_{\rm EQ}^2),
\end{equation}
 where
\begin{eqnarray}
 H^2 \equiv \left( \frac{\dot{a}}{a} \right)^2 = \frac{8 \pi }{3M_P^2} \rho ,
\end{eqnarray}
 with $n_{\rm EQ}$ and $a(t)$ being the equilibrium number density
 and the scale factor in a situation where the radiation dominates the universe with 
 the energy density $\rho=\rho_{\rm rad}$, i.e., the thermal equilibrium epoch. $\langle\sigma v\rangle$ is 
 the thermal averaged product of the annihilation cross section 
 by the relative velocity\cite{Bertone:2004pz}. As usually adopted,  we present our results in the form of $\Omega_N$, which is the ratio between the energy density of $N$ and the critical density of the universe. 

\begin{figure}[!ht]
\centering
\includegraphics[width=0.47\textwidth ]{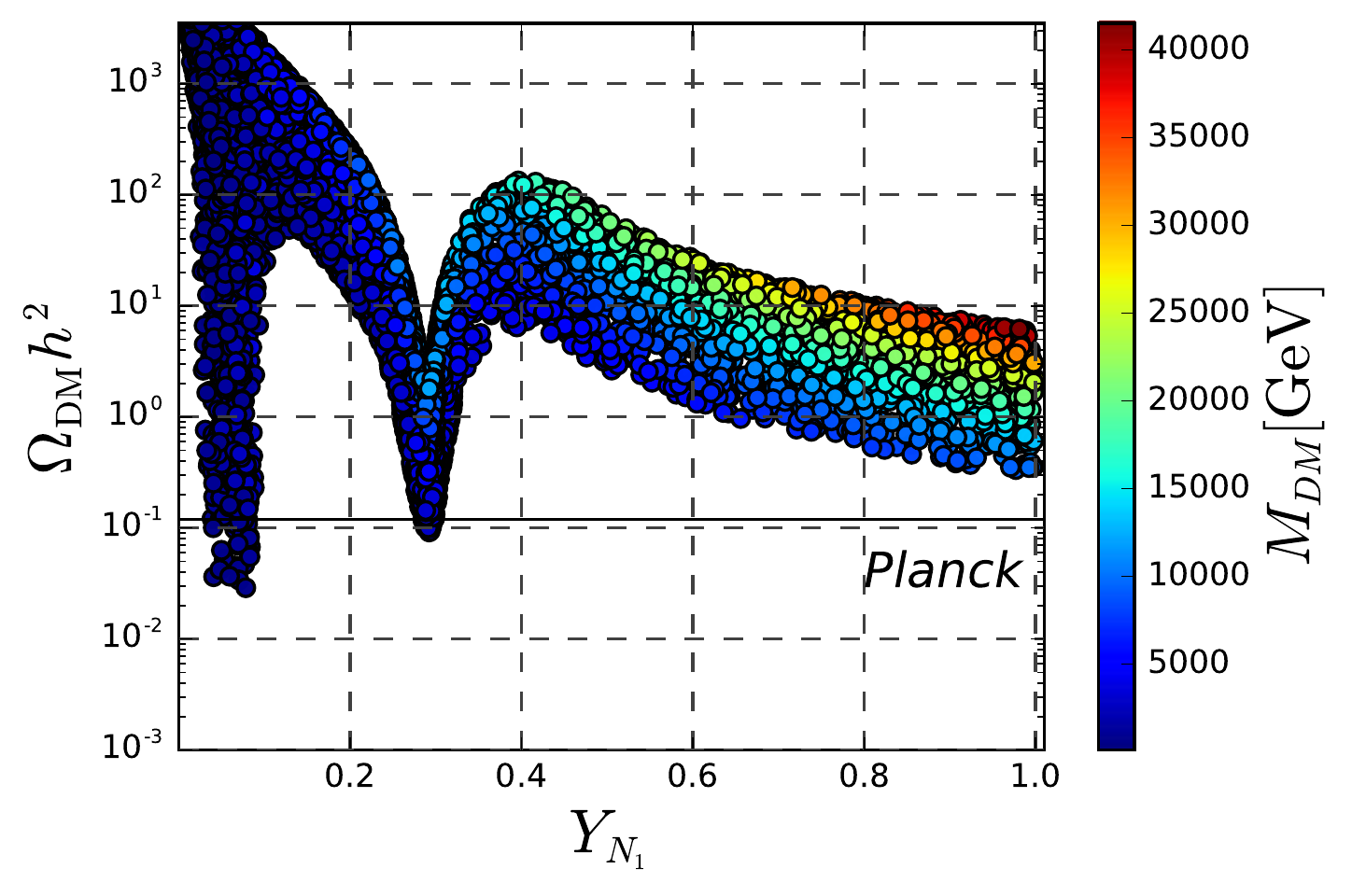}
\includegraphics[width=0.47\textwidth ]{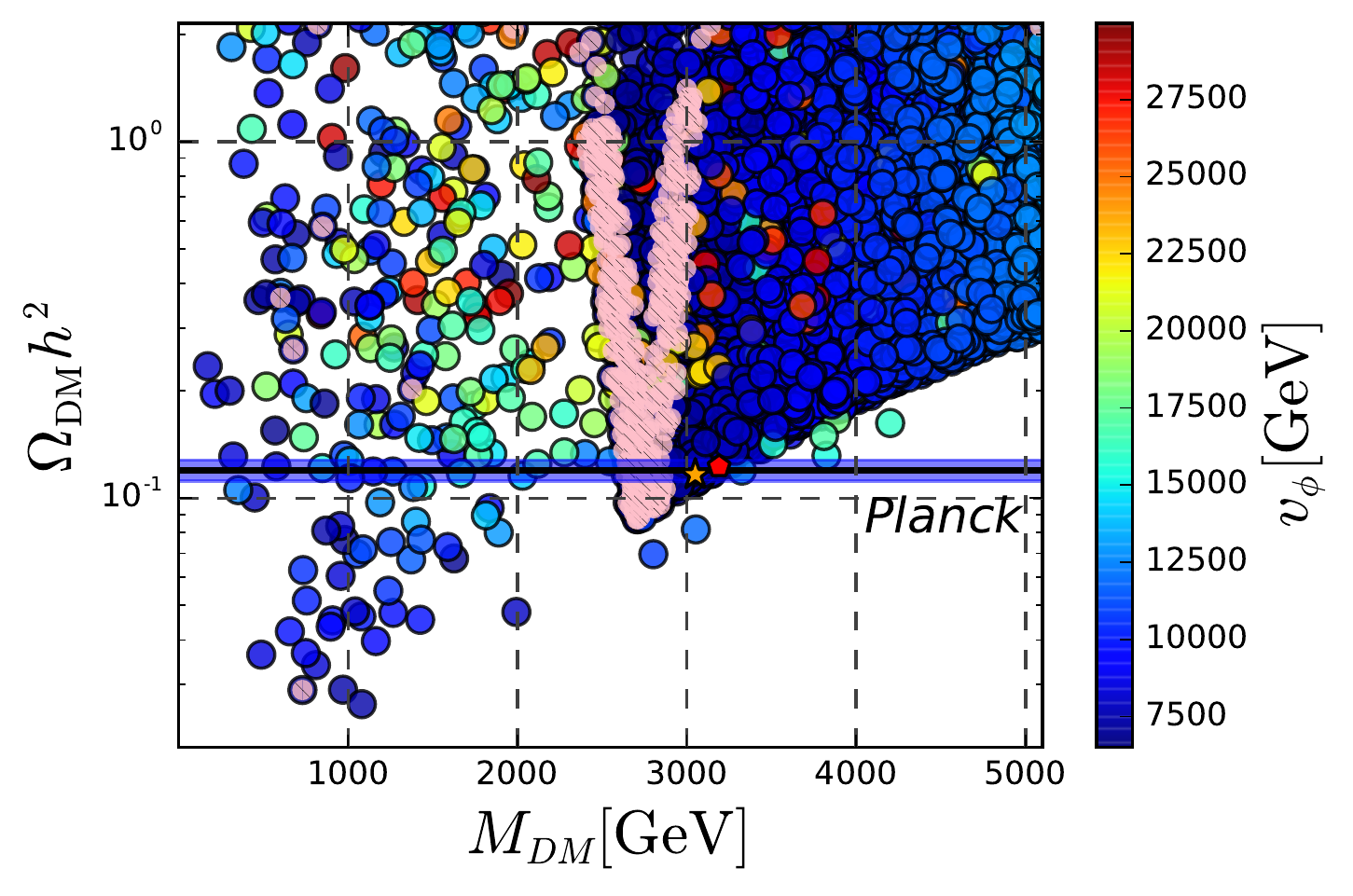}
\includegraphics[width=0.47\textwidth ]{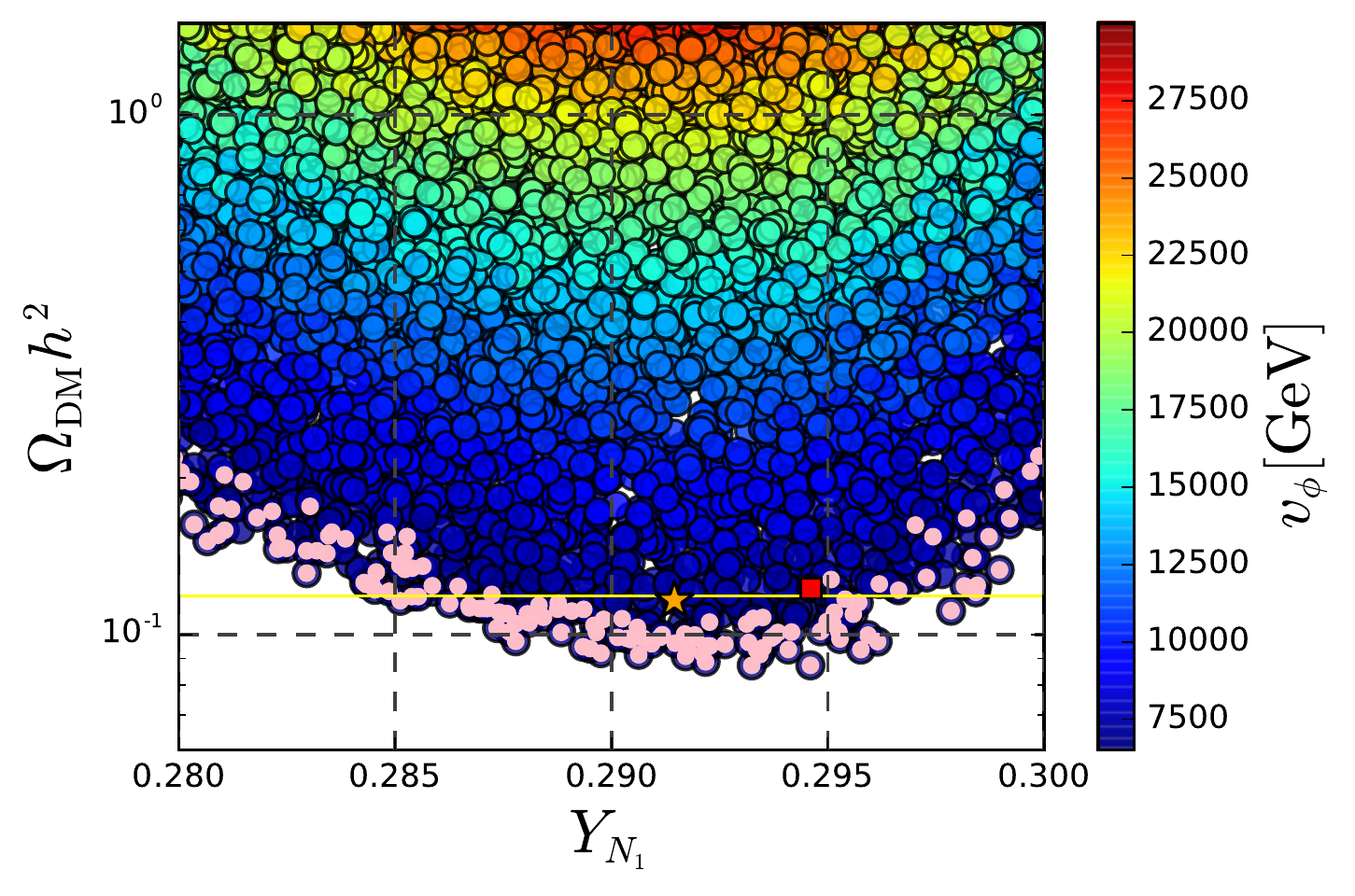}
\includegraphics[width=0.47\textwidth ]{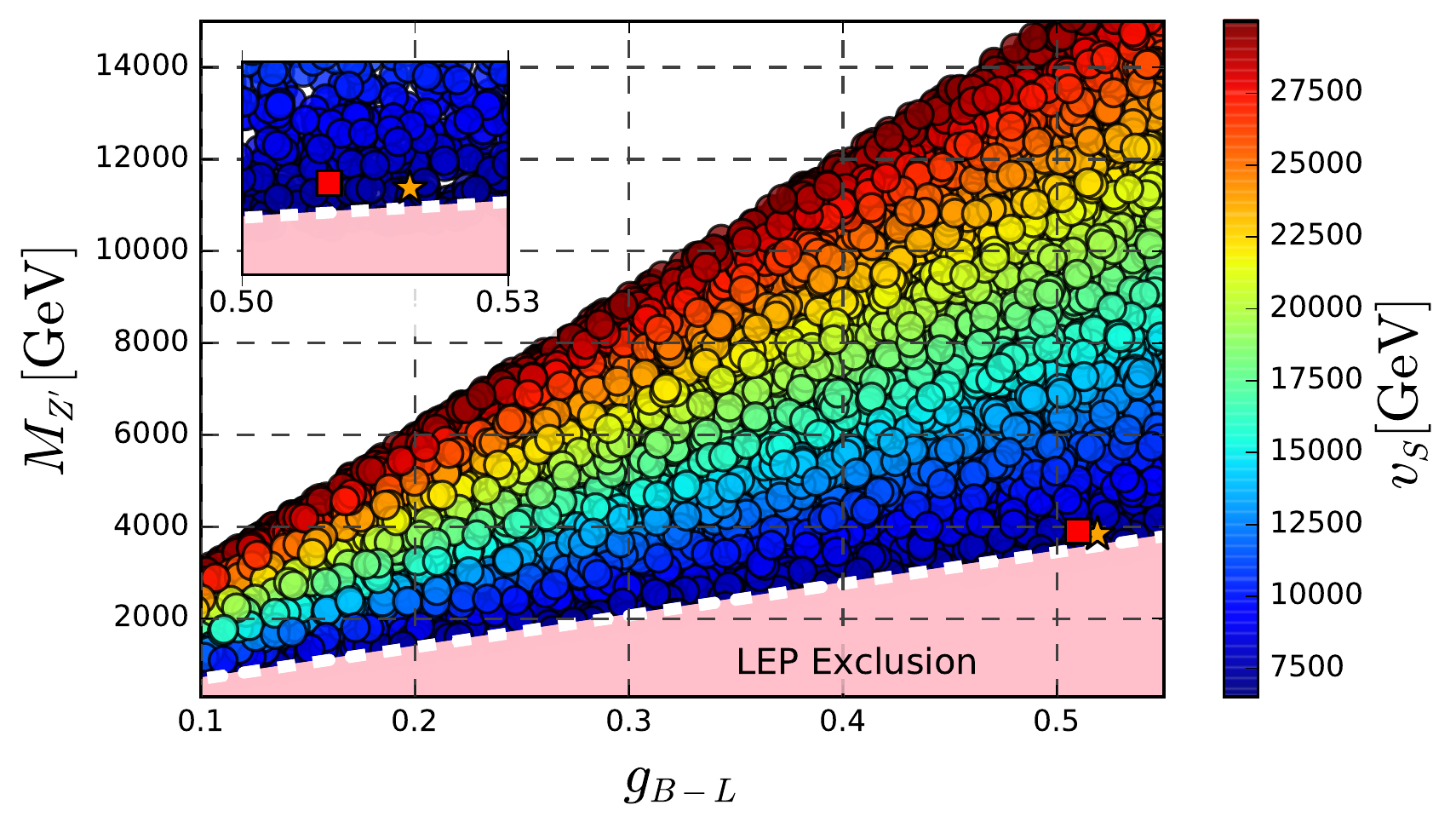}
\caption{ Plots relating DM relic Abundance, Yukawa coupling of the RHN and the  dark matter candidate mass. The thick horizontal line correspond to the correct relic abundance\cite{Patrignani:2016xqp}. }
\label{DMRelicY}
\end{figure}

We proceed as follow. We analyze  numerically the Boltzmann equation  by using the micrOMEGAs software package (v.2.4.5)\cite{Belanger:2010pz,Belanger:2010pz1}. For this we  implemented the model in Sarah (v.4.10.2) \cite{Staub:2008uz,Staub:2010jh,Staub:2012pb,Staub:2013tta} in combination with SPheno (v.3.3.8)\cite{Porod:2011nf,Porod:2003um} package,  which solves  all mass matrix numerically.  Our results for the relic abundance of $N$ are displayed in FIG. \ref{DMRelicY}. The thick line in those plots correspond to the correct abundance. Perceive that,  although the Yukawa coupling  $Y_{N}$ is an arbitrary free parameter, which translates in $N$ developing any mass value, however  what determine if the model provides the correct abundance of $N$ is the resonance of $Z^{\prime}$ and $H$. This is showed in the top left plot of FIG. \ref{DMRelicY}. In it the first resonance correspond to a $Z^{\prime}$ with mass around $3$ TeVs. The second resonance correspond to $H$ with mass in the range from $6$ TeV up to $7$ TeV. In the top right plot  we show the density of DM varying with its mass but now including the region excluded by LEP constraint given in Eq. (\ref{LEP}). On the bottom left  of FIG. \ref{DMRelicY} we made a zoom in the resonance of $H$. Perceive that we included the LEP constraint. For completeness reasons, on the bottom right  we show the dependence of $M_Z^{\prime}$ with $g_{B-L}$ including the LEP exclusion region, too. In the last three  plots we show two benchmark point localized exactly  in the line that gives the correct abundance. They are represented in red square and orange star points and their values are displayed in the table.  
Observe that LEP constraint  imposes $N$ reasonably massive with mass around few TeVs.  To complement these  plots, we show the ones in FIG. \ref{DMRelicYdiagrams1} which relates the resonance of $H$(all points in color) and the mass of the DM. The mixing parameter $\sin \theta$ is given in Eq. (\ref{eigenvectorsR}). Observe that LEP exclusion is a very stringent constraint  imposing $H$  with mass above $5800$ GeV and requiring DM with mass above $2800$ GeV. All those points in colors give the correct abundance but only those in black recover the standard Higgs with mass of $125$ GeV. The benchmark points in the red square and orange star are given in the table \ref{tablebench}. In summary, for the set of values for the parameters choose here $N$ with mass around 3 TeVs is a viable DM candidate once provides the correct relic abundance required by the experiments \cite{Ade:2015xua}. That is not all. A viable DM candidate must obey the current direct detection constraints.  
    \begin{table}[!h]
\begin{center}
  \begin{tabular}{ | c | c | c |c |c | c | c | c | c |}
    \hline
    $M_{DM}$(GeV) & $Y_{N1}$ & $M_{Z^\prime}$(GeV) & $g_{B-L}$ & $M_{H}$(GeV) & $ \Omega h^2 $ & $ v_S $   & $ \sigma_{DMq} $  & q  \\ \hline \hline
    3050 & 0.291 & 3840 & 0.518 & 6279 & 0.116 &   7400 &   5.4 10-11  & {\Large \textcolor{orange}{$\star$}}   \\ \hline
    3190 & 0.294 & 3904 & 0.509 & 6470 & 0.122 & 7658  & 5.4304e-11 & {{\footnotesize \textcolor{red}{$\blacksquare$}}}   \\ \hline
    \hline
  \end{tabular}
\end{center}
    \caption{Benchmark points for parameters values added on plots.}
    \label{tablebench}
    \end{table}
\begin{figure}[!ht]
\centering
\includegraphics[width=0.23\textwidth ]{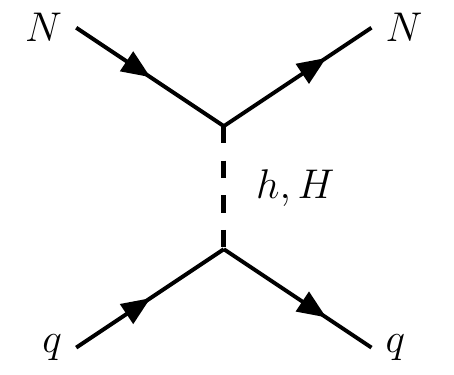}
\caption{The WIMP-quark scattering diagram for direct detection. }
\end{figure}
\begin{figure}[!h]
\label{DMRelicYmh2}
\centering
\includegraphics[width=0.7\textwidth ]{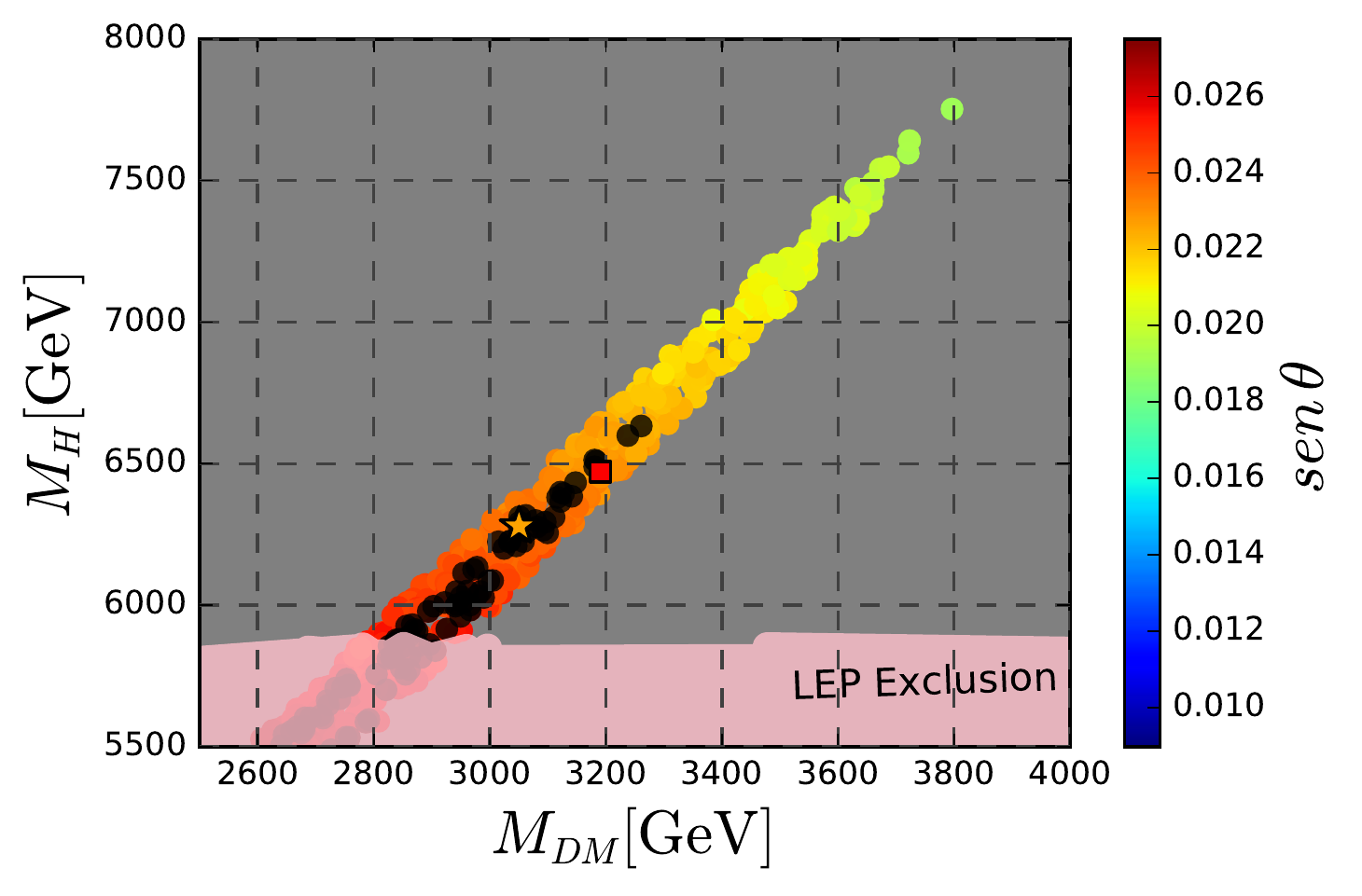}
\caption{ Physical parameter space of $M_{H_2}\times M_{DM}$. In colors we show the points that provide the correct relic abundance in the resonant production of the singlet scalar according with the diagrams in Fig. 4. The black points recover a Higgs with mass of $125$ GeV. }
\label{DMRelicYdiagrams1}
\end{figure}

In addition to the relic density of the DM candidate, which involves gravitational effects, only, we need to detect it directly in order to reveal its nature. Here we restrict our studies to direct detection, only. In attempting to detect directly  signal of DM in the form of WIMPs, many underground experiment, using different sort of targets, were built. Unfortunately no signal has been detected, yet. The results of such experiments translate in upper bounds into the WIMP-Nucleon scattering cross section. In view of this any DM candidate in the form of WIMPs  must be subjected to the current direct detection constraints. Direct detection theory and experiment is a very well developed topic in particle physics. For a review of the theoretical predictions for the direct detection of WIMPs in particle physics models, see \cite{Cerdeno:2010jj,Arcadi:2017kky,Queiroz:2017kxt}. For a review of the experiments, see \cite{Undagoitia:2015gya}. In our case direct detection requires interactions among  $N$ and quarks. This is achieved by exchange of $h$ and $H$ via t-channel as displayed in FIG. 4. Perceive that $Z^{\prime}$ t-channel gives  null contribution because $N$ is a Majorana particle. In practical terms we need to obtain the WIMP-quark scattering cross section which is given in \cite{Dutra:2015vca}. Observe that the scattering cross section is parametrized by four free parameters, namely $M_N$, $M_H$, $v_S$ and the mixing angle $\theta $ given in eq. (\ref{eigenvectorsR}). However this cross section depends indirectly on other parameters. For example, for $g_{B-L}$ in the range $0.1 - 0.55$ (as explained in Section \ref{sectionIII})  the LEP constraint implies $v_S > 7$ TeV and $M_{Z^{\prime}}$ around $3$ TeV. Considering this, and using the micrOMEGAs software package \cite{Belanger:2010pz,Belanger:2010pz1} in our calculations,  we present our results for the WIMP-Nucleon cross section in FIG. 6.  All color points conduct to the right abundance and are in accordance with the Lux (2017) exclusion bound. Those  points in pink are excluded by LEP constraint. However only those points in black recover a Higgs with mass of 125 GeV. Observe that the black points may be probed by future XenonNnT and DarkSide direct detection experiments. This turn our model a phenomenological viable DM model.
\begin{figure}[!ht]
\label{DMcrossq1}
\centering
\includegraphics[width=0.7\textwidth ]{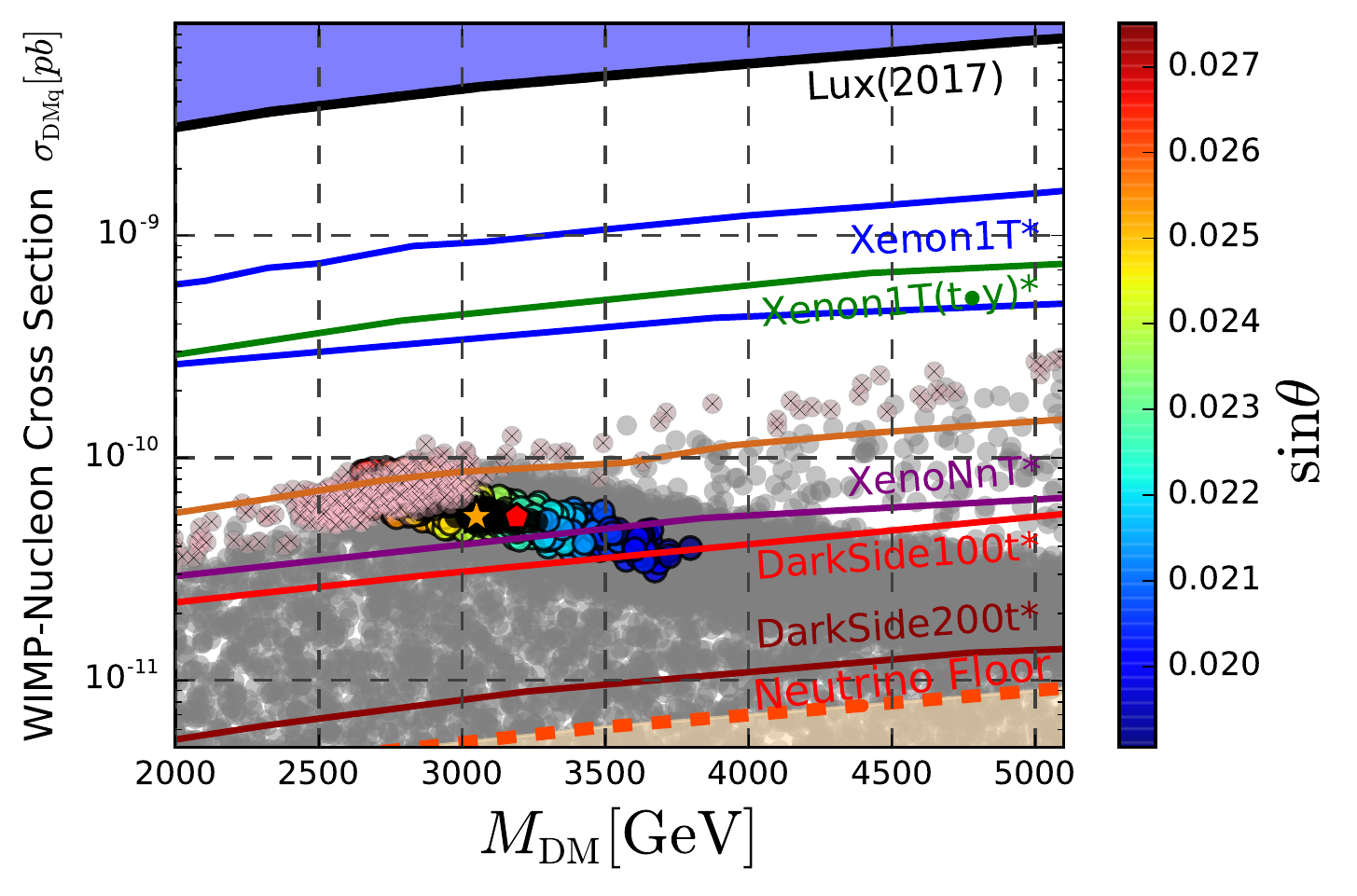}
\caption{ The spin-independent (SI) WIMP-nucleon cross sections constraints. The lines corresponds to experimental upper limits bounds on direct detection. 
  For LUX \cite{Akerib:2016vxi} (black line with blue fill area), Xenon1T
  \cite{Aprile:2018dbl} (blue line, prospect), XenoNnT \cite{Aprile:2015uzo} 
  (prospect, green line), Dark Side Prospect \cite{Aalseth:2017fik} (red and dark red lines for different exposure time) and the  neutrino coherent scattering, atmospheric neutrinos and diffuse supernova neutrinos \cite{Billard:2013qya} (orange dashed line with filled area).}
\end{figure}

\section{Conclusions}
For the first time in the literature the type II seesaw mechanism for generation of small neutrino masses was implemented within the framework of the B-L gauge model. We showed that neutrino masses at eV scale require  that $\Delta$ belongs to an energy mass scale around $10^9$ GeV. This characterizes a seesaw mechanism at intermediate energy scale and can be probed through rare lepton decays.

One interesting advantage of this model is that  we can evoke a $Z_2$ discrete symmetry and leave the right-handed neutrinos completely dark in relation  to the standard model interactions.  In this case these neutrinos turn out to be the natural candidate for the dark matter of universe in the form of WIMP. In this case we showed that the  correct abundance of dark matter is obtained thanks  to the resonant production of  $Z^{\prime}$ and of the heavy Higgs $H$.  Although our scenario is in accordance with LUX exclusion bound, however prospect direct detection experiments will be able to probe it.  

In what concern inflation, by allowing non-minimal coupling of the neutral  component of the scalar triplet with gravity, we showed that the model realize Higgs inflation in a very successful way since the model accommodates Planck results for inflationary parameters in a scenario where the loss of unitarity occurs orders of magnitude above the energy density during inflation. Furthermore, we obtained the prediction $Y_\nu(M_P) \simeq 0.460578$ for neutrinos Yukawa coupling at inflationary scale. As far as we know this is the first time an inflationary model gives such a precise prediction about neutrinos sector.


\section*{Acknowledgments}

The authors would like to thank Clarissa Siqueira and P. S. Rodrigues da Silva for helpful suggestions. This work was supported by Conselho Nacional de Pesquisa e Desenvolvimento Cient\'ifico - CNPq (C.A.S.P) and Coordena\c c\~ ao de Aperfeicoamento de Pessoal de N\'ivel Superior - CAPES (A.C.O.S and J.G.R).
\bibliography{bibliografia}
\end{document}